\documentclass[twocolumn,pre,ams]{revtex4}
\usepackage{graphicx}

\usepackage{epstopdf}
\usepackage{dcolumn}
\usepackage{amssymb,amsmath}
\usepackage{bm}

\begin{document}

\title{Geometrical Frustration in Two Dimensions:  Idealizations and Realizations of a Hard Disc Fluid in Negative Curvature}
\author{Carl D. Modes and Randall D. Kamien}  \affiliation{Department of Physics and Astronomy, University of Pennsylvania, Philadelphia, PA 19104-6396, USA}

\date{\today}

\begin{abstract}
We examine a simple hard disc fluid with no long range interactions on the two dimensional space of constant negative Gaussian curvature, the hyperbolic plane.  This geometry provides a natural mechanism by which global crystalline order is frustrated, allowing us to construct a tractable model of disordered monodisperse hard discs.  We extend free area theory and the virial expansion to this regime, deriving the equation of state for the system, and compare its predictions with simulation near an isostatic packing in the curved space.  Additionally, we investigate packing and dynamics on triply periodic, negatively curved surfaces with an eye toward real biological and polymeric systems.
\end{abstract}

\maketitle

\section{Introduction}

The true nature of glasses and the glass transition has long remained inscrutable.   Glasses belong to a wide class of disordered materials and systems which include granular materials like sand \cite{sand} or wheat grains \cite{wheat}, frustrated systems like spin glasses \cite{spin}, and soft systems like foams \cite{foam} and emulsions \cite{emulsions}.  Unfortunately, the analogies afforded by this classification have done little to lift the veil of mystery: from misunderstandings about apparent ``flow" displayed by stained glass windows from the middle ages \cite{stained} to debate over the meaning of randomness \cite{Torquato}, few aspects of the problem have escaped some measure of controversy.  The prominent role of geometrical frustration in these systems, however, has long been agreed upon \cite{geofrustr}, and it is here that we turn our attention and expand upon our prior work \cite{cdmrdk}.

We begin by considering an idealized colloidal glass, made of monodisperse spheres in three dimensions with no long-range interactions and a purely hard-core, short-range potential.  The entropically favored, high-density fcc lattice (there is a virtual tie between fcc and hcp lattices, and lattices with the stacking faults that relate them) for the system is well understood and highly stable, but, nevertheless the system can exhibit a glass transition to a dynamically arrested, amorphous state at a packing fraction significantly lower than that of the crystal.  One way to understand this is to recognize that there is a competition at play between local and global properties of the spheres' arrangements.  On the one hand, the best the colloids can hope to do, entropically speaking, is to arrange themselves into one of the close-packed, global crystal configurations.  On the other hand, these crystal configurations are \textit{not} the best local arrangements -- a tetrahedral packing is locally preferable, but cannot be maintained beyond small distances without the introduction of defects.

In spite of this qualitative understanding of the importance of geometrical frustration, the full colloidal glass system remains difficult to tame, particularly near the transition \cite{ohern}.  In principle, one could imagine constructing a simpler system that shares enough of the qualities of the colloidal glass that a full analysis yields insights into the more complicated case.  A time-honored tradition along these lines is the study of the same system in fewer dimensions \cite{lowd}.  In this case, we would have monodisperse, hard discs with no long-range interaction.  Notice, however, that we have lost geometrical frustration!  The entropically-favored, global, hexagonal crystal is composed precisely of the locally optimal units, triangular configurations of discs.  Indeed, two-dimensional, monodisperse discs do not exhibit a glass transition \cite{ohern}.

How then, can we use a simplified system to say anything about the genuine article?  We must find a way to modify the two dimensional case so as to recover some measure of geometrical frustration.  One method that has been well studied is the use of bi- or polydisperse discs \cite{ohern}, frequently with disc radii in a ratio of $1.4:1$.  The use of two or more disc sizes successfully reintroduces geometrical frustration and recovers the glass transition.  The drawbacks to this tack are the possibility of a phase transition into locally monodisperse, crystalline regions, and, the explicit introduction of additional particle types and the arbitrary nature of the size distribution, breaking the clean analogy with the original colloidal glass.

So, is there a way we can keep the two-dimensional system monodisperse and yet still manage to frustrate the local and global packings?  Indeed, in the seminal work of Kl\'{e}man, Nelson, and others \cite{NelsonPRL, geofrustr}, background curvature is introduced to the system, achieving the desired frustration.  To picture how this might happen, imagine that our two-dimensional space is a sphere, and that the hard discs are embedded in it as spherical caps.  Notice that there is no longer enough room around each spherical cap to fit six additional caps, as there would be in flat space.  More precisely, the number of discs (or spheres or hyperspheres) that can simultaneously touch a central one, the kissing number, $n_{\rm kiss}$, varies continuously as a function of the curvature, $K$.  In two dimensions, this relationship is given by \cite{nelson}:
\begin{equation}
n_{\rm kiss} = \frac{2\pi}{\cos^{-1} \left( \frac{\cosh (2\sqrt{-K}r)}{1 + \cosh (2\sqrt{-K}r)} \right)}.\label{kn}.
\end{equation}
In order for local triangular packings of discs to be commensurate with the best global packing, $n_{\rm kiss}$ must be an integer.  This condition is satisfied only for flat space ($K=0 \rightarrow n_{\rm kiss} = 6$), and for isolated non-vanishing curvatures.
 
Historically, curved backgrounds in this context have been used to study crystalline states on curved surfaces \cite{NelsonPRL, nelson}, and the ordered defect patterns that emerge as a result, or to imagine the ideal, unfrustrated, ordered configurations in three dimensions with the hope of understanding how ``decurving" might lead to the physical, disordered states we observe in glassy systems {\sl \`{a} la} the cholesteric blue phase \cite{geofrustr, blue}.  In contrast to these approaches, we choose to study the disordered fluid regime of hard-core systems in uniform curvature.  Positive curvature may, at first blush, seem more suitable due to its intuitive appeal and the fact that a sphere may be embedded in three dimensional Euclidean space ($\mathbb{R}^3$) without distortion.  However, since the surface area of the sphere is finite, and smaller particles on the sphere's surface probe ever lower curvatures (see Appendix A), there is no way to extract a sensible thermodynamic limit at a given curvature.

Instead, we turn to the space of constant negative curvature, the hyperbolic plane, $\mathbb{H}^2$, where the thermodynamic limit is not impossible, but remains subtle nonetheless.  While the sphere suffers from not having enough room for a thermodynamic limit, on $\mathbb{H}^2$ there may actually be \textit{too much} room.  To see why this might be so, consider a circle of radius $r$ on $\mathbb{H}^2$.  Its area is $A(r)=-2\pi K^{-1} \left[\cosh(\sqrt{-K}r) -1\right]\approx \pi r^2 -\frac{\pi}{12}Kr^4$ for small $r$, consistent with the known $K=0$ result.  Meanwhile, the circle's circumference is
$C(r)=2\pi  \sinh(\sqrt{-K}r)/\sqrt{-K}$, which grows as fast as $A(r)$ for large $r$.  Thus, the area within an arbitrary distance of the boundary of a circle, or any other fixed shape, is always a finite fraction of the entire region's area, no matter how large the region becomes.  Since the traditional thermodynamic limit depends on the effect of the boundary vanishing at infinite size, we must tread carefully.  Because our system has interactions that are purely local, the ideal gas provides a useful test case for managing the issue.

The textbook approach to the ideal gas law begins with
particle-in-a-box eigenenergies and takes the large area limit to calculate the 
partition function for a single particle, $Z_1=V/(\lambda_T)^d$ where $\lambda_T=\sqrt{2\pi\hbar^2/(m k_B T)}$ is
the thermal wavelength.   We might expect that the partition function, even on a curved manifold, would be proportional to the proper hypervolume of the box.  Indeed, since any partition sum is dominated by the infinite number of high energy, short wavelength terms, one would think that these modes are insensitive to the curvature as soon as their de Broglie wavelength is shorter than the radius of curvature.  For concreteness, let's consider the partition function for an ideal gas on a hemisphere where
the wavefunctions are forced to vanish on the equator; the energies are the eigenvalues of the Laplacian on the hemisphere with Dirichlet boundary conditions.  On the whole sphere, the spherical harmonics $Y_{\ell m}(\Omega)$ are the complete set of eigenfunctions with eigenvalue $\ell(\ell+1)$ for $m=-\ell,\ldots\ell$.
The eigenfunctions on the hemisphere will be a subset of the spherical harmonics, with the boundary condition enforcing the constraint that $\ell + \vert m\vert$ is odd and so the degeneracy for the $\ell(\ell+1)$ eigenvalue is $\ell$.  Thus the partition function will be:
\begin{eqnarray}
Z_1^D &=& \sum^{\infty}_{\ell=1} \ell \exp\left\{-\frac{\hbar^2 \ell(\ell+1)}{2 m k_B T R^{2}}\right\} \nonumber\\
&=& \sum^{\infty}_{\ell=1} \ell \exp\left\{-\frac{\lambda_T^{2} \ell(\ell+1)}{4\pi R^{2}}\right\}
\end{eqnarray}
where $R$ is the radius of the spere.  We can re-express this partition function as an expansion in $\lambda_T / R$ by employing the Euler-Maclaurin formula \cite{PadeEulerMac}:
\begin{eqnarray}
\sum_{n=i}^{j} f(n) 
&=& \int_{i}^{j} f(x) \, \, dx + \frac{f(i) + f(j)}{2} \\
&& + \sum_{k=1}^{\infty} \frac{b_{2k}}{(2k)!} \left( f^{(2k-1)}(j) - f^{(2k-1)}(i) \right) \nonumber
\end{eqnarray}
where $b_{2k}$ are the even Bernoulli numbers.  Substituting $f(n) = n\exp{\left[ \lambda_T^2 n (n+1) / 4 \pi R^2 \right]}$, $i = 0$ and $j = \infty$ gives the desired expansion:
\begin{equation}
Z_1^N = \frac{2\pi R^2}{\lambda_T^2} - \frac{\pi R}{2 \lambda_T} + \frac{1}{6} + {\cal O}(\lambda_T/R)
\end{equation}
Notice that we get terms proportional to both the area and the perimeter, but also a constant piece that one may speculate is related to the topology of the manifold.  Similarly, we could consider the partition function with Neumann boundary conditions along the equator, {\sl i.e.} vanishing derivative.  In this case the constraint is that $\ell+ \vert m\vert$ is even and the appropriate sum to consider is
\begin{eqnarray}
Z_1 &=& \sum^{\infty}_{l=1} (\ell+1) \exp\left\{-\frac{\hbar^2 \ell(\ell+1)}{2 m T R^{2}}\right\}\nonumber\\
&=&  \frac{2\pi R^2}{\lambda_T^2} + \frac{\pi R}{2 \lambda_T} + \frac{1}{6} + {\cal O}(\lambda_T/R)
\end{eqnarray}
Again we see an area and perimeter term, the latter with the opposite sign, and the same constant piece.  
We note that the sum of $Z_1=Z_1^D+Z_1^N$ is the partition function for the whole sphere and we see that 
the perimeter terms cancel leaving us with 
\begin{equation}
Z_1=\frac{4\pi R^2}{\lambda_T^2} + \frac{1}{3} + {\cal O}(\lambda_T/R)
\end{equation}
again, the area of the manifold plus an extra constant.  

To probe more exotic surfaces, one could attempt this kind of calculation on a case-by-case basis \cite{comment1}, but fortunately, the spectrum of the Laplacian on a general Riemann surface
is well characterized.  Originally, Weyl studied the asymptotic behavior of the large eigenvalues \cite{Weyl}.  Kac, in asking, ``Can one hear the shape of a drum?" \cite{Kac}, found our constant $\frac{1}{6}$ and,  finally,  McKean and Singer \cite{MS} developed the ``Weyl expansion'' for the partition function of a single particle on a two-dimensional domain $D$:
\begin{eqnarray}
Z_1(T) 
&=& \frac{1}{\lambda_T^2}\int_D dA \pm \frac{1}{4\lambda_T}\int_{\partial D} d\ell +\frac{1}{12\pi}\int_D K dA\nonumber\\
&&\quad +\frac{1}{12\pi}\int_{\partial D} \kappa_g d\ell +{\cal O}(\lambda_T)
\end{eqnarray}
where the sign of the first correction depends on Neumann (+) or Dirichlet (-) boundary conditions and $\kappa_g$ is the geodesic curvature of the boundary considered relative to its outward-directed normal \cite{comment2}.  Our calculations are, unsurprisingly, in complete agreement with their theorem which, interestingly, works in any dimension.  Returning to the question at hand, namely the hyperbolic plane,  we can see that the circumference directly enters the partition function at uncomfortably low order.  What is to be done?  Periodic boundary conditions provide an out -- by eliminating the boundaries and considering sufficiently large periodic boxes we may recover some semblance of a thermodynamic limit.  We hasten to point out that even this procedure is not free of its own subtleties -- due to the inherent length scale provided by the curvature (Appendix A), a simple re-scaling of the periodic box is not allowed.  In particular, by symmetry, the edges of the periodic box must have vanishing geodesic curvature.  Thus the Gauss-Bonnet Theorem which relates the background curvature $K$, the geodesic curvature $\kappa_g$ and the jump angles at any vertices, $\alpha_i$,
\begin{equation}
\int \!\!\! \int_{T} K \,dA = 2\pi - \sum_{i}\alpha_{i} - \int_{\partial T} \kappa_{g} \,ds
\end{equation}
imposes a relation between the area contained within the box and the polygon's angles.  At the same time, to preserve homogeneity after the periodic boundaries have been applied, any identified vertices must end up with $2\pi$ of angular space around them, thus fixing the box's area.  Topological considerations force the number of sides to be a multiple of 4 \cite{tarjus}, and we are left with the familiar square cell in flat space, along with a specific-size octagon, dodecagon, and so forth on $\mathbb{H}^2$.  Luckily, the area of these $p$-gons increases with $p$ \cite{tarjus}, and we are reassured that sufficiently large periodic boxes are achievable.

Recently, we took the first several steps in a comprehensive treatment of just such a system of hard discs on $\mathbb{H}^2$ \cite{cdmrdk}.  Our examination included a study of ordered disc packings leading to the application of free-area theory, a calculation of the first several virial coefficients of the system, and a molecular dynamics realization of the fluid in the smallest possible periodic region.  In this paper, we extend and detail these results by considering a constant-curvature recasting of the free-area theory and calculating the virial coefficients to higher precision while applying Pad\'{e} approximants to get a qualitative look at the behavior of the divergence in $\phi$ as a function of the curvature.  We expand our simulation to a larger polygon with more sides, allowing for better statistics through more simulated discs, and use it to probe the zero curvature limit from below.  In addition, we study entropically governed packings on surfaces with average negative curvature in $\mathbb{R}^3$, such as the triply periodic minimal surfaces found in diblock copolymers, lipids, or even the mitochondria of \textit{Chaos carolinensis} \cite{bingo}, among other biological examples.

This paper is organized as follows.  In section II, we review the theory of regular tesselations in two dimensions of arbitrary curvature, and extend this theory to packings of hard discs with crystalline order.  Section III develops a free area theory from these ordered packings on curved space.  In section IV, we calculate the first five terms in the virial expansion as a function of curvature, and comment on the analogy between high-dimensional flat space and the highly curved plane.  Section V presents a molecular dynamics model of a hard disc fluid on $\mathbb{H}^2$ and compares the simulation results with the predictions of free area theory and the virial expansion developed in earlier sections.  In section VI, we conisder a hard disc system constrained to live on a negatively curved triply periodic surface such as those found in the cell or among diblock copolymers.  We summarize our results in section VII.

\section{Tilings and Disc Packings}

We begin our study of the hyperbolic plane, and curved space in general, with a discussion of the classical problem of tiling a plane with regular patterns.  Regarding the hyperbolic plane in particular, we outline the features of our chosen model, the Poincar\'{e} disc model, in the Appendix.  

The problem of understanding the possible coverings of a flat plane with uniform tiles is deceptive in its simplicity, and the different shapes the tiles may take or patterns they may be adorned with have long been pondered.  In the last 100 years or so it has become known that there are precisely 17 ways of covering a flat plane with these tiles, corresponding to the 17 so-called Wallpaper Symmetry Groups \cite{wallpaper}.  The Alhambra, in Granada, Spain, for example, is famous for its artistic representations of all 17 possible tilings \cite{Alhambra}.

\subsection{Tesselating Two-Dimensional Spaces}

If we restrict ourselves to covering the plane with only \textit{unadorned} polygons, as one would consider in a Voronoi tesselation, however, we reduce the number of possible tilings considerably.  Furthermore, since crystalline order is a necessary starting point, we also restrict our tiles to regular polygons.  We then begin with a regular lattice of Voronoi cells and consider a tesselation of $F$ $p$-gons, with $q$ meeting at each vertex. We allow for the possibility that $F$ may be either finite, corresponding to a compact object, or infinite, corresponding to the Euclidean or hyperbolic plane.  Each $p$-gon has $p$ vertices, but each vertex is shared by $q$ equivalent $p$-gons.  Thus the total number of vertices is $pF/q$. Likewise, each edge is shared by two $p$-gons and so the total number of edges is $pF/2$. The total number of faces is $F$.  The Euler character $\chi$ of the surface constrains these three numbers and it follows that
 
\begin{equation}\label{euler}
\chi=V-E+F = pF\left[\frac{1}{ q}  + \frac{1}{p}- \frac{1}{2}\right].
\end{equation}

\begin{figure}[t]
\centering
\includegraphics[width=8.5cm]{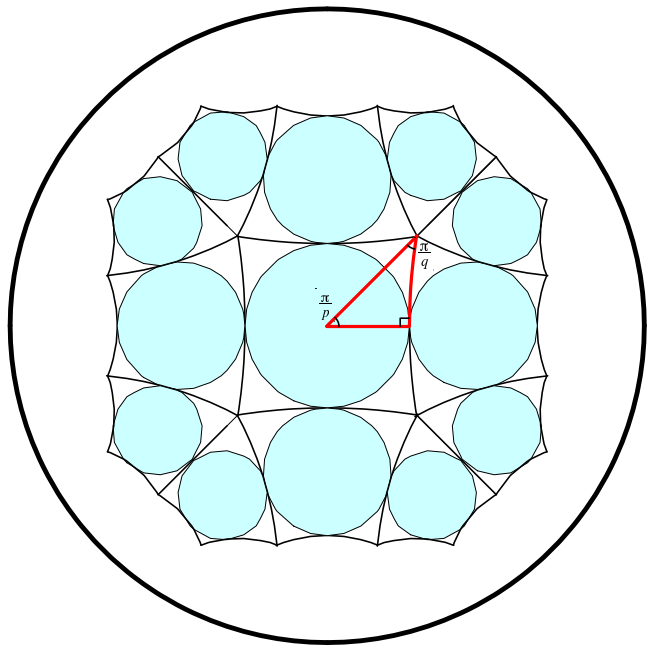}
\caption[4-5]{(Color online). A $\{4,5\}$ tiling of a segment of the hyperbolic plane, represented in the Poincar\'{e} disc model, with inscribed discs.  Despite the apparent differences, the discs are of uniform size and the tiles are all equivalent regular polygons.  The red triangle displayed in the central cell is used to calculate $\phi$. \label{4-5}}
\end{figure}

For the flat periodic plane (the torus) $\chi=0$, so $p^{-1}+q^{-1}=1/2$ and we find the three tilings represented by the Schl\"afli symbol $\{p,q\}=\{6,3\}$, $\{3,6\}$, and $\{4,4\}$, hexagons, triangles, and squares, respectively.  The spherical topology is the only one with positive 
$\chi (=2)$, admitting the five Platonic solids $\{p,q\}=\{3,3\}, \{3,4\},\{3,5\},\{4,3\}$, and $\{5,3\}$: the tetrahedron, octahedron, icosahedron, cube, and dodecahedron, respectively.  Finally, we turn to the pertinent geometries with negative Gaussian curvature, for
which $\chi<0$ and $p^{-1}+q^{-1} < \frac{1}{2}$.  In this case there are an infinite number of integral pairs $\{p,q\}$ and an infinite number of corresponding regular tesselations, for example, $\{4,5\}$ (Figure \ref{4-5}).  As in the tilings of the Euclidean plane, the $\{p,q\}$ tesselation is dual to
the $\{q,p\}$ tesselation -- the vertices of one become the face centers of the other.  However, this does not mean that their packing properties are the same, just as the packing of discs on the centers of hexagons is much more efficient than packing them on the centers of triangles.

\subsection{Building Ordered Disc Packings}

We use these tesselations to produce close-packed, crystalline-ordered arrangements of discs simply by inscribing circles into their constituent regular polygons.  Note that this procedure results in the familiar hexagonal crystal when applied to the $\{6,3\}$ tesselation of flat space.  In general, those packings produced from tesselations with $q=3$ will result in the most efficient, densest crystals, as $q=3$ corresponds to local triangular close-packing around each vertex.  The rigidity or stability of a packing is another feature of interest that may be inferred from the parameters of the packing.  In particular, a configuration is considered to be \textit{isostatic} \cite{maxwell} if there are exactly enough contacts to balance the degrees of freedom in the system, that is, the system is at the threshold of stability.  In an $N$-particle packing in $d$ dimensions there are $dN$ degrees of freedom and $\frac{1}{2} \bar{z} N$ contacts, where $\bar{z}$ is the average number of contacts per particle, so the condition for isostaticity is $\bar{z}=2d$.  Therefore, for an arrangement to be stable we must have $p \geq 4$, with isostaticity at $p=4$.  Note that isostaticity is a topological issue and is independent of the curvature of the surface.

A central property of these crystalline arrangements is their area fraction, as it indicates which packings are favored entropically near a given curvature.  We find the area fraction of the $\{p,q\}$-packing through application of the Gauss Bonnet Theorem and hyperbolic trigonometry.  When there are $F$ identical polygons, the area of each can be found from the Euler character $\chi$:
\begin{equation}
A_{\rm p-gon} = \frac{\int dA}{F} = \frac{-\int K dA}{K F} = \frac{-2\pi \chi}{KF} 
\end{equation}
The radius of the incircle can be determined via the dual law of hyperbolic cosines \cite{coxeter},  $\cosh(\sqrt{-K}r) = \cos(\pi/q)/\sin(\pi/p)$ (Figure \ref{4-5}).   The area fraction for the $\{p,q\}$-packing is thus:
\begin{equation}
\phi_{\{p,q\}}= \frac{\frac{\cos(\pi/q)}{\sin(\pi/p)} - 1}{p\left(\frac{1}{2} - \frac{1}{q} -\frac{1}{p}\right)}
\label{pqformula}
\end{equation}

We recover the appropriate packing fractions, $\pi/3\sqrt{3}$, $\pi/4$, and $\pi/2\sqrt{3}$, for the Euclidean tilings, $\{3,6\}$, $\{4,4\}$, and $\{6,3\}$, respectively.  As $p\rightarrow\infty$ for close-packed $q=3$ tesselations, we find $\phi = 0.9549$, the known packing fraction of the best packing on the hyperbolic plane at any curvature \cite{twelve}. The lowest packing fraction for an isostatic configuration ($p=4$) is $\phi_{\{4,\infty\}}=\sqrt{2}-1\approx 0.4142$, an area fraction far below that for the Euclidean square lattice $\pi/4\approx 0.785$.  Also of note is the broadening of the range in $\phi$ that supports stable configurations, from $\Delta\phi\sim 0.13$ in  two-dimensional flat space to $\Delta\phi\sim 0.54$ on  $\mathbb{H}^2$. 

\section{Free-Area Theory in a Uniformly Curved Background}

Now that we have explicitly constructed close-packed disc configurations with crystalline order, we have a foothold on the analysis of a thermodynamic fluid of hard discs living in a curved space.  Kirkwood's free-volume theory \cite{Kirkwood} is highly successful at modeling the equation of state for traditional crystals, particularly near their close-packed, maximum density \cite{Salzburg, Torqu}.  Somewhat surprisingly, free-volume theory remains successful at lower densities as well; as such, it provides a useful tool for probing a thermodynamic system, so long as the high density, high correlation behavior is known.

Before we apply free-volume theory to our curved system, recall that in flat space the shapes of the Voronoi cell and the free-volume cell are equivalent and, thus, only the scaling of the free-volume cell is important to the equation of state.  This scaling is dependent only on dimension, and we find a simple expression for the equation of state in terms of the maximum packing fraction $\phi_{\text{max}}$ and the hypervolume per particle, $v$ \cite{bookchapter}:

\begin{equation}
P_{\text{FV}} = -\frac{k_{B}T\phi^2}{v} \frac{d}{d\phi} \ln\{[(\phi_{\text{max}}/\phi)^{1/D} - 1]^{D}\}
\end{equation}
Unfortunately, an analogously simple treatment does not exist in the presence of a curved background.  The curvature itself, inducing geodesics to converge (positive $K$) or diverge (negative $K$), breaks the scale invariance enjoyed by all shapes on ${\mathbb{R}}^n$, and hence destroys the similarity relation between the Voronoi cell and the free-volume cell.

\subsection{An Idealized Circular Free Area Cell}

As an alternative, we consider an idealized case in which the symmetry of the system itself ensures that similarity is restored.  If, for example, we have a two-dimensional fluid composed of hard discs, and we posit that the free area cells are \em also \em circular, then we may proceed in direct analogy with Euclidean space.  The free area is simply the area of a disc whose radius is the difference between that of the Voronoi cell and that of the particle, $A_{\text{free}}=-\frac{2\pi}{K} (\cosh[\sqrt{-K}(r_{\text{cell}} - r_{\text{disc}})]-1)$.  The angle addition formula for the hyperbolic trigonometric functions, along with the recognition that $\phi$ is the ratio of $A_{\text{disc}}$ to $A_{\text{cell}}$, allows us to write:

\begin{eqnarray}
A_{\text{free}} &=& -\frac{2\pi}{K} \bigg{(} \frac{1}{\phi} (\cosh[\sqrt{-K}r] + \phi - 1)\cosh[\sqrt{-K}r] - \nonumber \\ 
& &\frac{1}{\phi}\sqrt{(\cosh[\sqrt{-K}r] - 1)(\cosh[\sqrt{-K}r] + 2\phi - 1)} \nonumber \\
& & \times \sinh[\sqrt{-K}r] - 1\bigg{)}.
\end{eqnarray}

\begin{figure}[t]
\centering
\includegraphics[width=8.5cm]{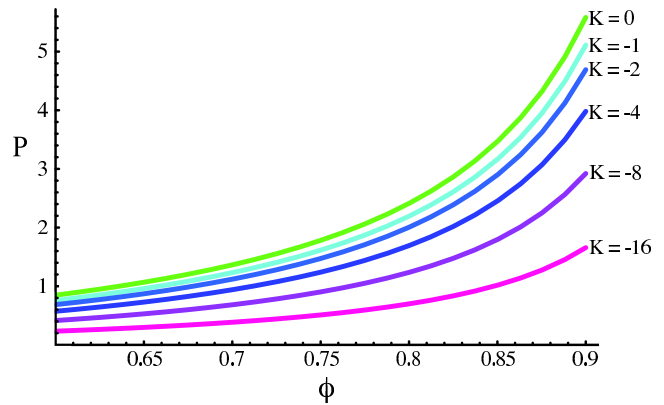}
\caption[disccell]{(Color online). The equation of state for free volume theory in a disc shaped Voronoi cell.  The top, green curve is for flat space; subsequent curves are at increasing negative curvature.  Warmer colors correspond to more curvature.  \label{disccell}}
\end{figure}

Notice that we recover the expected Euclidean result of $\pi r^2(1/\phi^{1/2}-1)^2$ as we take $K \rightarrow 0$.  Furthermore, denoting the proper area of the disc as $a$, we have

\begin{equation}
P_{\text{FA}} = -\frac{k_{B}T\phi^2}{a} \frac{d}{d\phi} \ln(A_{\text{free}})
\label{FAT}
\end{equation}
and it follows that the pressure diverges as a simple pole as $\phi \rightarrow 1$.  Notice, however, that as we increase the background curvature the divergence at $\phi = 1$ becomes sharper, apparent at ever higher area fractions  (see Figure \ref{disccell}), indicating a lower system pressure than is seen in a corresponding flat space system.  This suggests that both particle collisions and more complicated, many-particle arrangements become increasingly scarce as the curvature rises.  

\subsection{Free Area For Tesselation-Derived Packings}
\begin{figure}[t]
\centering
\includegraphics[width=8.5cm]{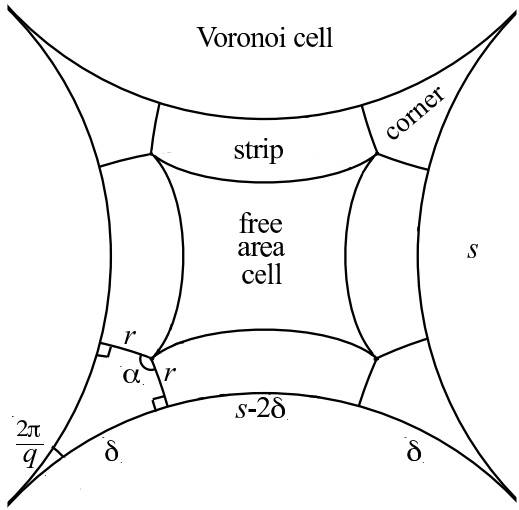}
\caption[pqFAT]{Partition of a $\{p,q\}$ Voronoi cell into its free area cell, $p$ corners, and $p$ strips.  Note that the boundaries of the free area cell are not geodesics.  \label{pqFAT}}
\end{figure}
\begin{figure}[t]
\centering
\includegraphics[width=8.5cm]{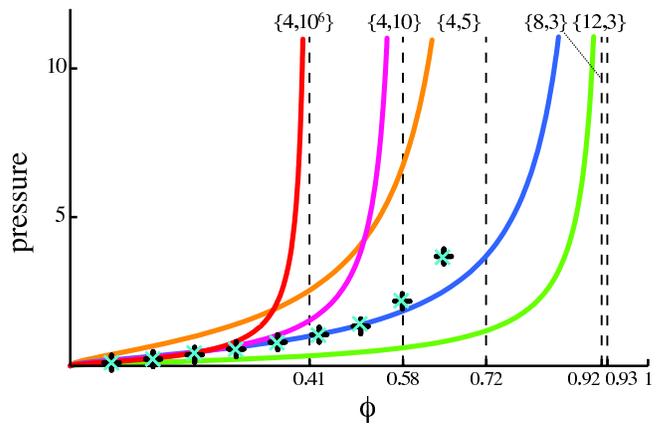}
\caption[pqpressure]{(Color online).  The equation of state for free area theory for a range of $\{p,q\}$ tesselations.  Note that the pressure curves diverge at their maximum packing fractions and that they differ even at low volume fractions, $\phi$. The vertical dashed lines indicate the maximum volume fraction for the tesselations listed on top of the figure.  The (light blue) Xs are the result of our molecular dynamics simulation for curvature just below the $\{4,5\}$ tesselation and the (black) crosses are for curvature just above $\{4,5\}$.  The number of particles ranges from 1 to 9.  \label{pqpressure}}
\end{figure}

What if we were to consider a non-idealized system based on the ordered packings discussed in section II, with polygonal Voronoi cells?  In order to cope with the shape distortions caused by the background curvature, we divide the Voronoi cell into individual regions, as depicted in Figure \ref{pqFAT}, and calculate the areas of the strip and corner regions separately, implying $A_{\text{free}} = A_{\text{Voronoi}} - pA_{\text{strip}} - pA_{\text{corner}}$.  Without loss of generality we may set $K=-1$ since
it is only the product of the disc size and $\sqrt{-K}$ which enter into all our expressions.  We can access different curvatures by changing the size of the discs. 
Direct application of the Gauss-Bonnet Theorem yields $A_{\text{Voronoi}} = p\pi - 2\pi -2p\pi/q$.  In order to find the other two areas, we first need $s$, $\delta$, and $\alpha$.  By considering the triangle formed by the Voronoi cell's center and two adjacent vertices we may apply the dual hyperbolic law of cosines (Appendix A.2) to find:

\begin{equation}
s = \cosh^{-1} \left[ \frac{\cos \left( \frac{2\pi}{p} \right) + \cos^{2} \left( \frac{\pi}{q} \right)}{\sin^{2}\left( \frac{\pi}{q} \right)} \right].
\end{equation}
To proceed we construct two identical triangles out of the corner $4$-gon by bisecting $\alpha$.  The dual hyperbolic law of cosines applied to the side of length $r$ then gives:

\begin{equation}
\alpha = \sin^{-1} \left[ \frac{ \cos \left( \frac{\pi}{q} \right)}{ \cosh (r)} \right].
\end{equation}
Armed with an expression for $\alpha$, we apply the hyperbolic law of cosines again, this time to the side of length $\delta$ yields:
\begin{equation}
\delta = \cosh^{-1} \left[ \frac{ \sqrt{ \cosh^{2} (r) - \cos^{2} \left( \frac{\pi}{q} \right) }}{\cosh(r) \sin \left( \frac{\pi}{q} \right) } \right].
\end{equation}
Now $A_{\text{strip}}$ must be the area within a given distance (namely, $r$) of a geodesic of length $s-2\delta$.  This area is $(s - 2\delta) \sinh(r)$ (see Appendix A.3).  Given $\alpha$, direct application of the Gauss-Bonnet Theorem gives $A_{\text{corner}}$.  Finally, we may eliminate $r$ through $\phi$:

\begin{equation}
r = \cosh^{-1} \left[ \frac{p \phi}{2} - \frac{p \phi}{q} - \phi +1 \right]
\label{elim}
\end{equation}
Collecting terms and setting $\tau = 1 - \frac{\chi \phi}{F}$ and $\tau_{pq}=1-\frac{\chi\phi_{\{p,q\}}}{F}=\cos(\pi/q)/\sin(\pi/p)$, we have

\begin{eqnarray} 
A_{\text{free}}  &= & 2p \sin^{-1} \left[\cos\left(\frac{\pi}{q}\right) \frac{\tau_{pq}}{\tau} \right]- 2 \pi\nonumber\\ &&- 2p \sqrt{\tau^{2} - 1}
\cosh^{-1} \left[ \frac{\cos(\frac{\pi}{p})}{\sin(\frac{\pi}{q})} \right]  \\
& & +2p \sqrt{\tau^{2} - 1} \cosh^{-1} \left[ \frac{\sqrt{1-\frac{\tau_{pq}^2}{\tau^{2}}\sin^2(\frac{\pi}{p})}}{\sin(\frac{\pi}{q})} \right] \nonumber
\end{eqnarray}
As a check of the complex algebra, note that as $\phi\rightarrow\phi_{\{p,q\}}$, $\tau\rightarrow \tau_{\{p,q\}}$ and $A_{\text{free}}\rightarrow 0$.  Likewise, as $\phi\rightarrow 0$, $\tau\rightarrow 1$ and 
\begin{equation}
A_{\text{free}} \sim 2p\left(\frac{\pi}{2}-\frac{\pi}{q}\right)-2\pi =-2\pi p \chi/F = A_{\text{Voronoi}}
\end{equation}
Unfortunately, away from these limits, this expression is difficult to analyze, so
we instead turn to Figure \ref{pqpressure}, where we plot the free-area equations of state for various $\{p,q\}$ tesselations using Equation (\ref{FAT})

\begin{figure}[t]
\centering
\includegraphics[width=8.5cm]{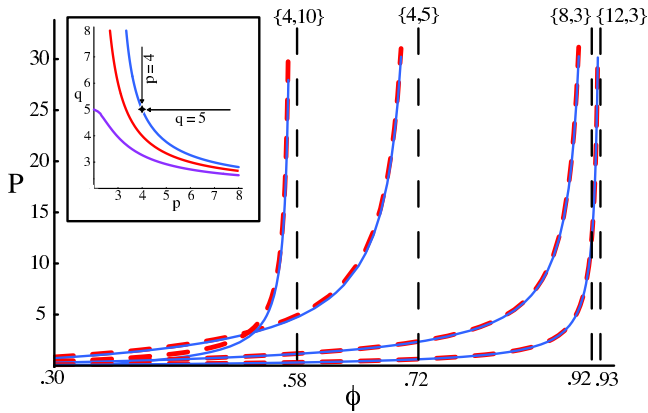}
\caption[pqspace]{(Color online).  Fixed radius equations of state with a sampling of $\{p,q\}$ tesselations as their divergent endpoints.  The solid blue curves are for fixed $q$, and the dashed red curves are for fixed $p$.  Note that in almost all cases these curves lie directly on top of one another.  The inset displays the parameter space for the packing parameters $\{p,q\}$.  The thick red line is the allowed combinations of $p$ and $q$ in flat space and the purple line beneath it a representative example from positively curved space.  The blue curve gives the allowed combinations for $\mathbb{H}^2$ with $r=0.53$.  The $\{4,5\}$ packing is identified and the paths through parameter space taken by holding either $p$ or $q$ constant are shown.  \label{pqspace}}
\end{figure}

It is important to note that by fixing the lattice of Voronoi cells we are forced to change the packing fraction by changing the disc size.  Since this is equivalent to changing the curvature, our free-area equation of state samples different curvatures at different area fractions $\phi$.  If we wish to apply this free-area theory to a system of constant disc size to sample a constant background curvature, we must allow the Voronoi scaffolding to vary continuously.   We will have to consider tesselations with non-integral $p$ and $q$; we can interpret these as the average number of edges per polygon and the average number of polygons meeting at a corner in a tesselation made of a variety of polygons. This is in the spirit of recent work on average polyhedra \cite{Plateau} which have a non-integral number of sides.  To maintain the curvature as we vary $\phi$, we might fix $p$ or $q$ adjusting the other so as to keep $r$ fixed in (\ref{elim}).  We might also hold $p/q$ fixed and set $p$ to maintain $r$.  
The freedom afforded by choosing one's path through $p$-$q$ parameter space is troubling -- there seems to be no {\sl a priori} way for one such path to be chosen over another, and we are forced to wonder how the behavior of the system is determined.  Fortunately, upon considering specific cases, the problem disappears, as the fixed-radius equation of state displays insensitivity to this choice (Figure \ref{pqspace}), particularly near high density, indicating that perhaps there is a hidden, unaccounted-for symmetry among the isotropic Voronoi scaffoldings.    We only find path dependence for very large values of $q$ and $p$ -- there are no real solutions for $p$ when $q$ is large until we get close
to the divergence in the pressure, though there the divergence is insensitive to the details again.

\section{The Virial Expansion to the Equation of State}

Thus far, by considering the extension of free-area theory to uniformly curved spaces we have been working in the high-density regime, since free-area theory is the first order term in an expansion about a close-packed state.  On the other hand, we are also interested in the system's behavior when it is much more sparse.  The expansion about low densities provided by the virial expansion to the equation of state \cite{landau} is suitable in this regime, although the connection between the two regimes is still not well understood in hard-core systems \cite{dohard?}.  In contrast to free-area theory, the virial expansion's validity is dependent on the ability to take a proper thermodynamic limit.  We therefore resort to the periodic spaces discussed earlier as our solution to this difficulty.

We note, however, that for hard discs, the virial expansion only includes local interactions among clusters of particles.  In fact, for pure, hard-core potentials, the successive terms of the virial expansion have a simple intuitive interpretation.  The first order term, as always, is simply the ideal gas law.  The next term removes the contribution from particle pair configurations that are disallowed by the steric interaction, that is, pairs of particles whose hard-cores overlap.  But this over-counts, as the same configuration with three or more particles simultaneously overlapping is subtracted multiple times.  The third virial term explicitly corrects this problem for three-particle arrangements, but similarly over-counts for larger numbers of particles, this time in the other direction.  Higher terms continue to both correct over-counting errors from previous terms and contribute smaller over-counting errors of their own, in a manner reminiscent of an infinite tower of Venn diagrams.  Fortunately, as long as the local clusters under consideration do not wrap around the periodic space and self-interact, we may ignore the topology introduced by our usage of a periodic space and perform the cluster integrals on ${\mathbb{H}}^2$.  In order to ensure that this is possible, we may continue to increase the area available to a cluster by constructing ever higher genus manifolds \cite{tarjus}.  Furthermore, for manifolds with constant $K<0$, $Z_1=A[\lambda_T^{-2} +K/(12\pi)]>0$ when 
$\lambda_T^2<-12\pi/K$.  In other words, there is an ample range of $K$ in which to work where the curvature of the manifold does not probe the quantum regime of the gas.  Hence, the pressure
$P=Nk_{B}T d\ln Z/dA$ is independent of the proportionality constant.  

Just as in the last section, we take advantage of the lack of scale invariance on ${\mathbb{H}}^2$ to probe higher curvatures not by changing the ambient space, but simply by considering larger and larger discs.  We consider discs of radius $r\sqrt{-K}$.  Without loss of generality we again fix the Gaussian curvature to be $K = -1$.  Dimensional analysis may always be employed to insert the appropriate factors of $K$. 
As in flat space, the 
second Virial coefficient, $B_2(r)$, is given by half the excluded area of a single disc.  To see that this is true, recall that, in general, $B_2(r)$ is given by
\begin{equation}
B_2(r) =  -\frac{1}{2V} \int \!\!\! \int f_{12} \,d^2{\bf r_1} \,d^2{\bf r_2}
\end{equation}
where $f_{12}$ is the Mayer function and the position integrals for both particle 1 and particle 2 are carried out over all space.  Since ${\mathbb{H}}^2$ is isotropic and homogeneous, we may integrate out one particle position and pick up a factor of $V$.  The remaining integration is over the separation between the discs.  Meanwhile, the Mayer function for hard discs in the Poincar\'{e} disc model is:
\begin{equation}
f_{ij} = \left\{ \begin{array}{ll}
		-1 & \mbox{if $\cosh^{-1}\left[1+2\frac{x^2}{1-x^2}\right] < 2r$}\\
		0       & \mbox{otherwise}.\end{array} \right.
\end{equation}
so long as one of the discs is at the origin. The expression when both discs are away from the origin is more complicated but follows from (\ref{distance}).  However, for the purposes of $B_2(r)$, we can
always put one disc at the origin.
Turning the step function into a condition on the bounds of the integral over $x$, inserting the metric from the Poincar\'{e} disc model (see Appendix A), and integrating out the angular part results in the following expression for $B_2(r)$:
\begin{equation}
B_2(r)=4\pi \int^{\sqrt{\frac{\cosh(2r) - 1}{\cosh(2r) + 1}}}_0 \,dx \,\frac{x}{(1-x^2)^2}.
\label{B2}
\end{equation}
Notice that this is precisely the circularly-symmetric, hyperbolic, area integral needed to calculate the area of a disc of radius $2r$ on ${\mathbb{H}}^2$, and indeed, carrying out the integration leaves us with $B_2(r) = \pi\left[\cosh(2r) - 1\right]$, which is half the excluded area of a single disc, the classic result from the virial expansion.
Further note that as $r\rightarrow 0$, $B_2\sim 2\pi r^2$, we recover the $d=2$ Euclidean result; in $d$ dimensions
$B_2(r) = (4\pi)^\frac{d}{2} r^d/ [\Gamma(\frac{d}{2}) d]$.

We may carry out a similar procedure to calculate higher coefficients.  The general relation for $B_3(r)$ is:
\begin{equation}
B_3(r) = -\frac{1}{3V} \int \!\!\! \int \!\!\! \int f_{12} f_{23} f_{13} \,d^2{\bf r_1} \, d^2{\bf r_2} \, d^2{\bf r_3}
\end{equation}
and we may still integrate out one of the disc positions.  Unfortunately, regardless of which disc we set to the origin, we are now guaranteed to retain a Mayer function where neither of the discs whose interaction it represents have had their position integrated out.  This means that we are forced to contend with an integral of the form
\begin{eqnarray}
B_3 &=& -\frac{32\pi}{3} \int_0^b \!\!dx_1\!\!\int_0^{2\pi}\!\! d\theta_{12}\!\!\int_0^b \!\!dx_2\\ 
&&\quad\left[\frac{x_1}{(1-x_1^2)^2} \frac{x_2}{(1-x_2^2)^2} f_{12}(x_1,x_2,\theta_{12}) \right]\nonumber
\end{eqnarray}
where $b$ is the same upper limit as in Eq. \ref{B2}, and the remaining Mayer function is written in terms of the coordinate distance of the two particles from the origin and the angle between them:
\begin{equation}
f_{12} = \left\{ \begin{array}{ll}
		-1&  \mbox{$\left[1+2\frac{\displaystyle x_1^2 + x_2^2 - 2x_1x_2\cos\theta_{12}}{\displaystyle\left(1-x_i^2\right)\left(1-x_j^2\right)}\right] < \cosh(2r)$}\\
		0       & \mbox{otherwise}.\end{array} \right. \nonumber
\end{equation}

\begin{figure}[t]
\centering
\includegraphics[width=8.5cm]{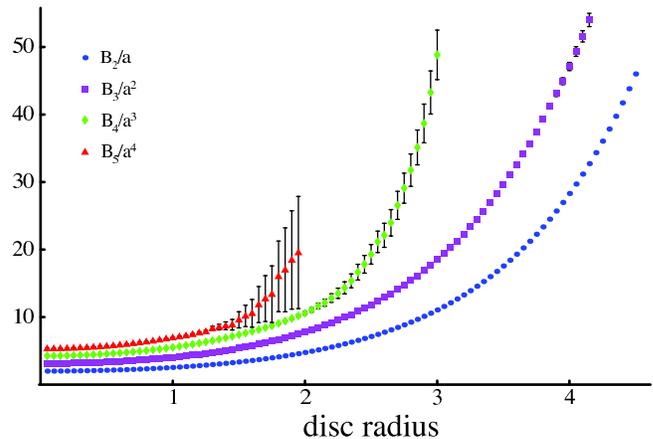}
\caption[B3B4B5plot]{(Color online).  The second, third, fourth and fifth virial coefficients in appropriate units of the area, $a$, as functions of disc size (for $K =-1$) $r$.  The zero size limit matches the flat-space coefficients.  $B_2$ is exact and error bars are indicated for all other coefficients.  These errors are given by the estimated variance of the mean of the integrand, and, for large $\eta$, scale like $\eta^{-1/2}$, where $\eta$ is the number of Monte Carlo trials. \label{B3B4B5plot}}
\end{figure}

In order to evaluate this integral, and the similar ones that appear in the higher virial coefficients, we turn to numerical, Monte Carlo techniques.  In addition, as is traditional in flat space \cite{lubanbaram}, the higher coefficients $B_n$ are reported in units of $a^{n-1}=\left[A(r)\right]^{n-1}$ to suppress size effects and generate the expansion for
$P/(\rho k_{B}T)$ in the area fraction, $\phi=\rho a$:
\begin{equation}\label{phivirial}
P=\rho k_{B} T\left[1 + \sum_{n=2}^\infty \frac{B_n}{a^{n-1}}\phi^{n-1}\right]
\end{equation}  
We have checked the $r\rightarrow 0$ limit in all cases, and find agreement with the known Euclidean results.
We plot $B_2/a$, $B_3/a^2$, $B_4/a^3$, and $B_5/a^4$ in Figure \ref{B3B4B5plot} as functions of $r$.    To the accuracy of our calculation, over the range of curvatures considered, $B_3$, $B_4$, and $B_5$ all remain positive.  There is evidence that, in higher dimensional flat spaces, the virial coefficients alternate in sign \cite{lubanbaram, virialD} and, hence, the leading singularity that controls the radius of convergence of the expansion sits on the negative real axis \cite{virialD}.  The location of this singularity, and thus the range of applicability of the virial expansion itself, remains an open question in $D = 2,3$. Though not conclusive, our results here suggest that the introduction of curvature does not \textit{qualitatively} change the behavior or location of the leading singularity in the expansion. 

\subsection{Pad\'{e} Approximants and the Radius of Convergence}

The quantitative behavior of the virial expansion's divergence as a function of the curvature, on the other hand, may be extracted from the coefficients that we have calculated, at least as a trend.  To this end we have approximated the virial series by ratios of rational functions, the Pad\'{e} approximants \cite{PadeEulerMac}.  We find the coefficients of two polynomials $a_m(\rho)$ and $b_n(\rho)$, of degree $m$ and $n$, respectively so that
\begin{equation}\label{pade}
P=k_BT\left(\rho+\sum_{j=2}^5 B_j\rho^j\right) = \frac{a_m(\rho)}{b_n(\rho)}
\end{equation}
The left hand side of (\ref{pade}) is fifth order and so there are, in principle, six arbitrary coefficients.  As a result, we can only study the Pad\'e approximants with $n+m\le 5$.   We look for the divergences in these approximants to estimate where the virial expansion diverges -- an estimate of the ultimate packing fraction if the virial expansion converges for all physical $\rho$.  Our results are shown in Table \ref{Pade}.
Notice that, for the Pad\'{e} approximants with only one pole (that is, of the form $[m,1]$) there is a clear trend towards divergence at lower $\rho$ as the curvature increases.  Furthermore, it is worth pointing out that among these approximants, the one that is the most sensitive, the $[3,1]$, also agrees best with the putative location of the divergence of the fluid branch in flat space, at around $\phi \approx 0.84$, as $r$ goes to zero.

\begin{table}[h]
\begin{tabular}{|c|c|c|c|c|c|}
\hline
 & \multicolumn{5}{c|}{$\phi^{*}_{R_{[m,n]}}$} \\
 \cline{2-6}
$[m,n]$ & $r \rightarrow 0$ & $r=0.5$ & $r=1.0$ & $r=1.5$ & $r=2.0$ \\
\hline
$[0,4]$ & -3.444 & 4.335 & --- & --- & ---\\
$[1,3]$ & 2.720 & -0.592 & -8.695 & 0.797 & 0.650\\
$[2,2]$ & --- & --- & --- & 0.798 & 0.695\\
$[3,1]$ & 0.798 & 0.797 & 0.794 & 0.770 & 0.676\\
$[0,3]$ & -5.145 & -45.09 & 0.634 & 0.409 & 0.258\\
$[1,2]$ & --- & --- & 0.888 & 0.821 & 0.829\\
$[2,1]$ & 0.735 & 0.733 & 0.732 & 0.730 & 0.737\\
$[0,2]$ & 0.737 & --- & --- & --- & ---\\
$[1,1]$ & 0.639 & 0.637 & 0.630 & 0.620 & 0.611\\
$[0,1]$ & 0.500 & 0.470 & 0.393 & 0.298 & 0.210\\
\hline
\end{tabular}
\caption{The leading divergences on the real axis in $\phi$ at various values of curvature and for different Pad\'{e} approximants about the origin.  Empty entries indicate that all poles for that approximant are located off the real axis.  Though $\phi$ has physical meaning only in the range $[0,1]$,   a divergence at an unphysical value of $\phi$ may still be meaningful, if, for example, it encodes information about the radius of convergence of the virial expansion.}
\label{Pade}
\end{table}

\section{Molecular Dynamics on the Hyperbolic Plane}

Equipped with these theoretical models for the system, particularly in the high and low density regimes, we turn to the full-blown hard-disc fluid.  We choose to employ constant energy molecular dynamics.  There is some subtlety in handling the boundaries of the simulation region: recall that the area near the edge of a shape scales at the same rate of the area of the entire shape on $\mathbb{H}^2$.  This means that the effect of hard boundaries cannot be minimized by considering larger regions.  We thus choose to implement periodic boundary conditions, although there is subtlety here as well.  As mentioned above, because of the scale dependence of the hyperbolic plane and the topological requirements of the periodic region, we are restricted to choosing for our simulation region the unit cell from tesselations of the form $\{4n,4n\}$ for $n$ an integer greater than $1$.  This ensures both that the sides of the polygon may be identified in a systematic way to produce an orientable manifold and that, upon identification, there is not a conical singularity at the vertices of the polygon (so there are exactly $2\pi$ radians of space around each vertex).

We have chosen not to use the event-driven molecular dynamics traditional for pure hard core systems, as projecting the next collision is extremely time intensive when the equations of motion are nontrivial as they are in a curved background.  Instead, we evolve the system by time steps, checking for and resolving collisions as they occur.  The collisions themselves require some care as well.  The center of mass frame in a curved space is not necessarily inertial, even if the constituent particles are all following geodesics -- imagine a sphere with a particle moving around the equator and another sitting motionless on the North pole -- the center of mass follows the 45$^{\hbox{th}}$ parallel which is not a geodesic.  As a result, the textbook techniques and formulae for handling collisions do not apply.  As an alternative, we decompose each particle's velocity into the component along the geodesic connecting the particle centers at contact and the component normal to this geodesic.  Note that these directions will differ between the two colliding particles in the local co\"ordinate system  due to the complexity of parallel transport on $\mathbb{H}^2$.  In order to resolve the collision, the components of velocity along the contact geodesic are swapped, and the other components remain unchanged.  Hence, collisions are guaranteed to conserve energy and momentum.

To calculate the equation of state from our simulation, we employ the collisional virial expression for the pressure \cite{collisional}:
\begin{equation}
P = \frac{1}{A} \left( NkT + \frac{1}{2 \tau} \sum_{\text{collisions}} d |\Delta {\bf p}| \right)
\end{equation}
where $A$ is the area of the simulation region, $\tau$ is the time over which collisions are summed, $d$ is the disc diameter and $\Delta {\bf p}$ is the momentum transferred from one particle to another during the collision.  Finally, we remind the reader that to fill out the equation of state with different values of $\phi$ it is necessary to change the number of particles simulated, $N$, as varying the disc size means changing the effective curvature.

\subsection{Systems Near the $\{4,5\}$ Tesselation}

We probe curvatures $r=0.53$ and $r=0.531$, just {\sl below} and just {\sl above} that for the $\{4,5\}$ tesselation, $r_{\{4,5\}}=0.5306$.  At these curvatures no regular tesselation is allowed, and so non-crystalline arrangements give the best packings \cite{radin}.  Thus we expect to probe disordered configurations even at high densities with this choice.  At lower densities, however, the gas will still be sensitive to the $\{4,5\}$ tesselation and should follow the free-area theory prediction for the $\{4,5\}$, failing only at higher area fractions.  In particular, we expect that the system's pressure should be lower than that of the isostatic crystal for higher $\phi$ and diverge at larger $\phi$.  This is precisely the case in three-dimensional flat space where the liquid branch is lower in pressure than the the free-volume crystalline pressure at low volume fractions.    Indeed, these expectations are borne out by the data (Figure \ref{pqpressure}).

\subsection{Approaching Flat Space from Below: Monodispersity and Disorder}

In addition to investigating the system at packing fractions close to an isostatic crystal, we might use a limiting procedure to take the background curvature to zero.  By looking at the fluid in low curvatures, we expect to gain some insight into hard discs in flat space.  In particular, looking at curvatures near, but not at, zero guarantees that we will avoid crystallization, as no crystal packings are available in this curvature range.  It is therefore possible that we may generate disordered, monodisperse configurations that can be carried into flat space with low probability of introducing overlaps or defects.  Unfortunately, due to the broken scale invariance on $\mathbb{H}^2$, a very large number of simulated discs is needed to probe this regime, making the simulation numerically untenable, even for the high-speed digital computers of our day. 

\section{Packing and Dynamics on Negatively Curved Surfaces in Flat Space}

Thus far, we have used uniform, negative, background curvature as a way of constructing an idealized model in two dimensions for a more complicated three dimensional system.  Owing to the fact that no surface with uniform, negative curvature can be isometrically embedded into ${\mathbb{R}}^3$, however, this model is unphysical.  But are there examples of real systems that are qualitatively close, for which the predictions of our model may be brought to bear directly?  Triply periodic minimal surfaces, with their numerous physically advantageous properties \cite{PRSA}, their prevalence in smectic liquid crystals, diblock copolymers, and even biology, and their status as infinite surfaces with negative curvature everywhere are ideal for analysis with insights from the abstract hyperbolic plane.

In lieu of a direct molecular dynamics simulation, as done on ${\mathbb{H}}^2$, we instead take advantage of the numerical insight provided by examining allowed and disallowed configurations \cite{KamienLiu}.  In fact, we may obtain the pressure simply by randomly sampling particle configurations at a given density and determining the fraction that violate the no-overlap condition of the hard-core potential.  If the probability of not finding an allowed state at density $\phi + \delta\phi$, given an allowed state at $\phi$, is $J(\phi)\delta\phi$, then the following relation holds \cite{KamienLiu}:
\begin{equation}\label{KL}
J(\phi) = \frac{V}{\phi k_BT}[P(\phi) - \phi T/v]
\end{equation}
where $v$ is the area per disc, $V$ the system volume, $T$ the temperature.

Hence, this approach gives us the equation of state.  We have traded the numerical difficulty of dynamics of particles constrained to a complicated surface with the algorithmic difficulty of finding a way to generate points at random on an arbitrary surface such that the distribution is flat with respect to the local area.

\subsection{Randomly Distributed Points on an Arbitrary Surface}

We have employed a novel geometric algorithm for selecting points from an arbitrary surface randomly with even distribution in the area (so long as it is either compact or periodic in each dimension).  Knowledge of the metric is not required, only a set of parametric equations that define the surface in $\mathbb{R}^3$.  The algorithm works as follows.  First, one repeat unit of the surface in whichever directions are periodic, and the entire surface in those directions for which the surface is compact, are considered to be inside a box.  At this stage it is necessary to determine the maximum number of times a straight line can intersect this region of the surface, say $I_{\text{max}}$ -- this can typically be done analytically, though for a very complicated surface numerics may prove more efficient.  The box is inscribed in a sphere, and a great disc is chosen at random in this sphere.  Next, a point is chosen at random on this great disc and the line normal to the great disc is drawn through it.  Finally, we determine the intersections of this line with the surface and randomly select one from among them.  To ensure that the resulting distribution of points is flat with respect to area, it is necessary to begin the entire process anew if the line fails to intersect the surface.  Furthermore, if the number of intersections is less than $I_{\text{max}}$ then an appropriate number of placeholders must be included when randomly selecting an intersection point.  If a placeholder is selected then the entire procedure is repeated.  If an actual intersection point is selected, then that point is the output of the pass through the algorithm.

To prove that this procedure yields a distribution flat with respect to the local area, consider the shadow cast by the surface when light from an arbitrary direction strikes it.  Now consider the contribution to this shadow from a differential piece of the surface.  The area of the part of the shadow cast by this small patch will be directly proportional to the likelihood of selecting a point from it when the great disc chosen at random is normal to the direction of light, $P(dA) \propto dA\vert\cos(\theta)\vert$, where $\theta$ is the angle between the patch's surface normal and the light rays.  But the algorithm specifies that we constantly choose, at random, new directions for the illumination.  This has the effect of averaging over all incoming directions for the light, $\theta$ and $\phi$, and removes the dependence on the angle, leaving the likelihood of selecting a point from any given patch of surface directly proportional to only the area of the patch.  To see this, imagine a particular great disc parameterized by $x$ and $y$.  The surface has a shadow on the disc of area $A_{\text{shadow}}$.  Ignoring, for the moment, multiple leaves we may parameterize the surface in terms of $x$ and $y$ so that the probability of selecting any particular area element $dA$ given that we pick a point in the shadow is:
\begin{eqnarray}
P(dA &\vert& \text{the point is in the shadow}) = \\
&& = \frac{dA\vert\cos\theta\vert}{\int_{\text{shadow}} dx dy} = \frac{dA\vert\cos\theta\vert}{A_{\text{shadow}}} \nonumber
\end{eqnarray} 
Note that $A_{\text{shadow}}$ depends on the choice of great disc, so this conditional probability
does not pick area elements with equal weight.  However,
the probability of picking a point in the shadow is just $A_{\text{shadow}}/A_{\text{great disc}}$.  Putting this together we see that $P(dA) =dA\vert\cos\theta\vert/A_{\text{great disc}}$.  Integrating over all directions yields a flat distribution.  Finally, if there are multiple leaves of the surface behind one another with respect to a given direction of incident light, randomly selecting from among them and multiple placeholders, as described above, guarantees that no patch is penalized or rewarded for its position relative to distant pieces of the surface.  

\subsection{The Plumber's Nightmare}

The simplest triply periodic minimal surface, and the one that we consider first, is the Schwarz P-Surface, sometimes referred to as ``The Plumber's Nightmare" for the impression it gives of a never-ending maze of pipe work.  This surface appears in the study of lipid bi-layers \cite{Plipids} and liquid crystals \cite{Pliqcry}.  In lieu of using the exact minimal surface, however, we use the trigonometric approximation, as its departure from minimality affects only the mean curvature, leaving the Gaussian curvature nearly unchanged.  Our approximate Schwarz P-Surface is defined by the following relation:
\begin{equation}
\cos(x) + \cos(y) +\cos(z) = 0.
\end{equation} 

\begin{figure}[t]
\centering
\includegraphics[width=8.5cm]{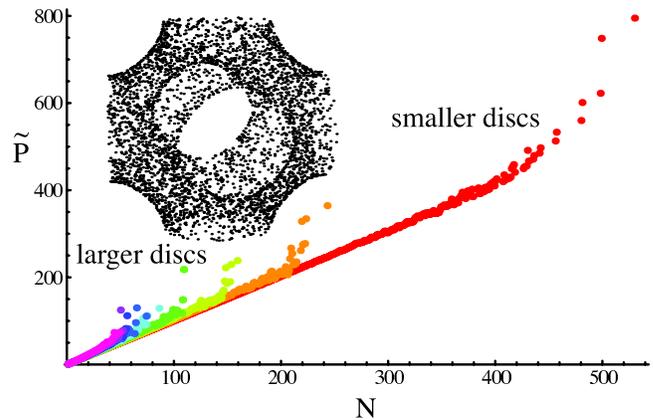}
\caption[PPress]{(Color online).  Dimensionless pressure as a function of $N$ for discs of various sizes living in the P-Surface.  The disc diameters range from 0.05 to 0.5 in units where the repeat length of the P-Surface cell is $2\pi$.  Warmer colors correspond to smaller discs.  A subset of the random points used is shown in the upper left corner.\label{PPress}}
\end{figure}

Using this level set approximation, we have used our algorithm to choose points as depicted in Figure \ref{PPress}.  We have checked our method by projecting these points along an axis and normalizing by the square root of the metric to see a uniform density.  With this in hand we may now calculate the pressure using (\ref{KL}).  We lay down $N$ discs of a given size, one at a time, checking to see if they can be inserted without overlapping an existing disc.  To accept or reject each disc, we check the distance between points along the surface and compare this to the disc diameter.  The distances between disc centers along the surface are also estimated in a metric-free way, by taking a small step along the line in $\mathbb{R}^3$ connecting the disc centers and finding the plane normal to this line at the location of the step.  We then calculate the curve of intersection between the P-Surface and the plane and find the point on this curve closest to the initial disc center.   Tabulating this distance and then repeating the process with the location of the last step as the new initial point from which the next line in $\mathbb{R}^3$ is drawn completes the algorithm.  Armed with an unambiguous state acceptance criterion, we have an avenue to calculate $J(\phi)$, however, our relevant control parameter is $N$.  So we instead write:
\begin{equation}
J(N) = \frac{V^2}{NTv}\left[ P(N) - NT/V\right]
\end{equation}
using $\phi = Nv/V$.  Isolating a dimensionless pressure, $\tilde{P} = P(N)V/T$ allows us to wash out the hard core system's insensitivity to the temperature and the overall scale invariance, leaving:
\begin{equation}
\tilde{P} = N\left( 1 + \frac{v}{V}J(N) \right).
\end{equation}
The dimensionless pressure calculated in this way is shown in Figure \ref{PPress} as a function of $N$.  Notice that for all considered disc sizes, the pressure shows little deviation from an ideal gas state $(\tilde{P} = N)$ until near the divergence.  This stands in excellent agreement with our predictions from free-area theory and meshes well with the observations made on hard core systems in general by Kamien and Liu \cite{KamienLiu}.

\subsection{\textit{Chaos Carolinensis} and the Schwarz D-Surface}

Another triply periodic minimal surface of interest is the Schwarz D-Surface, or Diamond Surface.  Lipid bi-layer and liquid-crystalline systems also both exhibit this surface, though it appears in a more surprising context as well -- biology.  When the amoeba \textit{Chaos Carolinensis} is deprived of food, its mitochondrial membrane transitions from an amorphous, unpatterned state to a D-Surface, possibly to optimize its surface-to-volume ratio while cannibalizing its own membrane phospholipids for energy \cite{bingo}.  We again choose to use the trigonometric approximation; this time the defining relation is given by:
\begin{equation}
\cos(x-y)\cos(z) + \sin(x+y)\sin(z) = 0.
\end{equation}

\begin{figure}[t]
\centering
\includegraphics[width=8.5cm]{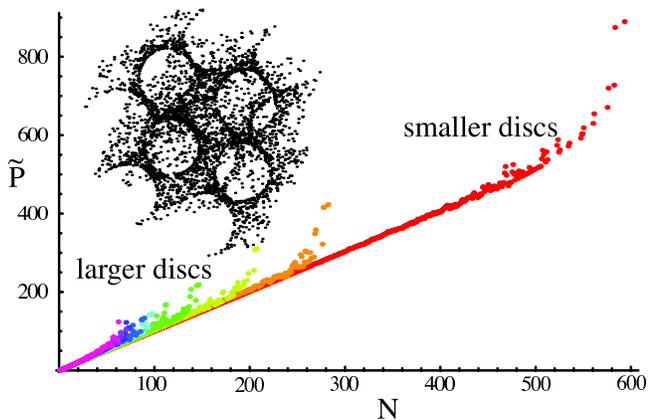}
\caption[DPress]{(Color online).  Dimensionless pressure as a function of $N$ for discs of various sizes living in the D-Surface.  The disc diameters range from 0.05 to 0.5 in units where the repeat length of the D-Surface cell is $2\pi$.  Warmer colors correspond to smaller discs.  A subset of the random points used is shown in the upper left corner.\label{DPress}}
\end{figure}

As before, we generate points on this surface (see Figure \ref{DPress}), and again we calculate the dimensionless pressure.  For the full range of considered disc sizes there is little deviation from the ideal gas until near the divergence, similar to the P-Surface, as shown in Figure \ref{DPress}.  Perhaps this is not surprising, as the fundamental cell of the P- and D-Surfaces are made of the same local pieces patched together in different ways.  Thus the purely local interactions are likely insensitive to this longer range structure leaving open the possibility that only interactions on the scale of the fundamental cells might be sensitive to the changed topology.

\section{Discussion}

In summary, we have constructed a monodisperse, two-dimensional system with hard-core repulsions by introducing negative background curvature to frustrate local close-packing.  This suppresses crystallization and brings the qualitative features of the system more in-line with the frustrated three-dimensional flat case than with the un-frustrated two-dimensional system.  We presented theoretical models of the system both at low and high density, with the Virial expansion and the enumeration of crystalline states along with a free-area theory based around these crystal endpoints.  We simulated the system in a periodic piece of $\mathbb{H}^2$, and compared these results with our theory and with the well-known flat-space systems.  Finally we numerically analyzed a similar system on a physically realizable curved surface with non-constant negative curvature.

We emphasize, however, that these are only the first steps into a rich geometric framework.  Technological advances will allow a clean look at the system as the curvature is turned off and will make higher Virial coefficients attainable.  More complicated particle shapes and interactions will allow a complete modeling of sub-micron scale packings constrained to curved membranes, as exist in lipids, diblock copolymers, and abound in biology.  In addition, it is worth considering $\mathbb{H}^2$ and the delicate balancing of factors that allow the thermodynamic limit to be trivially sensible in flat space and difficult to grasp anywhere else.

\begin{acknowledgements}

It is a pleasure to acknowledge discussions with N. Clisby, C. Epstein, O. L. Halt, A. Liu, M. L\'{o}pez de Haro, B. McCoy, C.D. Santangelo, Y. Snir, and D. Vernon.  This work was supported in part by NSF Grant DMR05-47230, a gift from L.J. Bernstein, and a gift from H.H. Coburn.  RDK appreciates the hospitality of the Aspen Center for Physics where some of this work was done.

\end{acknowledgements}

\appendix
\section{Classical Non-Euclidean Geometry}

Here we review some properties of $\mathbb{H}^2$ and the well-known Poincar\'{e} model, and collect some useful classical identities and theorems on uniform curved geometry in general \cite{coxeter,hypgeom}.

\subsection{The Poincar\'{e} Disk Model: Metric and Distance}

We employ the Poincar\'{e} disc model of $\mathbb{H}^2$, popularized by Escher \cite{escher}, which conformally maps the hyperbolic plane  with curvature $K$ to the disc of radius $\sqrt{-K^{-1}}$ on $\mathbb{R}^2$.  Using the coordinates $(x,y)$, the metric is \cite{tarjus}

\begin{equation}\label{metric}
ds^2 = \frac{4\left(dx^2+dy^2\right)}{\left[1 +K\left(x^2+y^2\right)\right]^{2}}
\end{equation}
where $x^2+y^2\le -1/K$. 

Geodesics in this model are arcs of circles which intersect the bounding circle normally;  circles in the hyperbolic plane remain circles in this model, but their coordinate centers and radii vary, shifting outward from the apparent center and shrinking far from the origin, respectively.
The radius of curvature sets a lengthscale and, consequently, any system on $\mathbb{H}^2$ is manifestly not scale invariant. 

In addition, it is useful to understand that the Poincar\'{e} disc model arises via stereographic projection \cite{hypgeom} in the same way as spherical mappings, and, in particular, maps of the earth.  Define the following quadratic form: $Q({\bf x}) = -x^2 - y^2 +z^2$.  Then the collection of points with norm one with respect to this quadratic form define a two-sheeted hyperboloid.  We carry out the stereographic projection by mapping points on the sheet at positive $z$ onto the unit disc in the $x$-$y$ plane via the collection of lines passing through the point $(0,0,-1)$.  Given a point in the unit disc, $(u_x,u_y)$ we now have the map to the hyperboloid:
\begin{equation}
(u_x,u_y) \rightarrow \left( \frac{2u_x}{1-u^2} , \frac{2u_y}{1-u^2} , \frac{1+u^2}{1-u^2} \right)
\end{equation}
where $u$ is the length of the vector in the $x$-$y$ plane.  The distance function for the Poincar\'{e} disc model is then given by the symmetric bilinear form related to $Q$ by polarization:
\begin{equation}
\cosh(d({\bf \tilde{u}}, {\bf \tilde{v}})) = \langle {\bf u}, {\bf v} \rangle = \frac{1}{2}(Q({\bf u} + {\bf v}) - Q({\bf u}) - Q({\bf v}))
\end{equation}
where we have denoted the two-dimensional vectors in the disc by ${\bf \tilde{u}}$ and ${\bf \tilde{v}}$ and their images on the hyperboloid in $\mathbb{R}^3$ by ${\bf u}$ and ${\bf v}$.  Substituting the expressions for the images in terms of the model vectors and simplifying yields the distance function for the Poincar\'{e} disc model with $K = -1$:
\begin{equation}
d(\mathbf{\tilde u},\mathbf{\tilde v}) =  \cosh^{-1} \left[ 1 + 2 \frac{| \mathbf{\tilde u} - \mathbf{\tilde v} |^{2}}{(1 - \tilde u^{2})(1 - \tilde v^{2})} \right].
\label{distance}
\end{equation}
Were we to write ${\bf \tilde u}={\bf \tilde v} + d{\bf \tilde u}$ then we find
\begin{eqnarray}
ds^2&=&\left[d({\bf \tilde u},{\bf \tilde v}) \right]^2= \left(\cosh^{-1}\left[1+2\frac{d{\bf \tilde u}^2}{(1-\tilde u^2)(1-\tilde v^2)}\right]\right)^2\nonumber\\
&\approx& \frac{4 d{\bf\tilde u}^2}{(1-\tilde u^2)^2}
\end{eqnarray}
in agreement with (\ref{metric}).

\subsection{The Hyperbolic Law of Cosines}

Consider a triangle in the hyperbolic plane with side lengths $a, b, c$ and corresponding angles $\alpha, \beta, \gamma$.  Call the coordinate vertices $A, B$, and $C$.  As before, we set $K = -1$ in order to simplify notation.  Now, because $\mathbb{H}^2$ is isotropic and homogeneous, we may assume that vertex $A$ sits at the origin of the Poincar\'{e} disc model.  The distance function (\ref{distance}) then gives the following relations:
\begin{equation}
C = \sqrt{\frac{\cosh(b) - 1}{\cosh(b) + 1}} \, \, \, \, \text{and} \, \, \, \, B = \sqrt{\frac{\cosh(c) - 1}{\cosh(c) + 1}}.
\label{BandC}
\end{equation}
At the same time, applying the distance function in a similar way to the side of the triangle without a vertex on the origin gives:
\begin{equation}
\cosh(a) = 1 + 2\frac{|{\bf B} - {\bf C}|^2}{(1-B^2)(1-C^2)}.
\label{fin}
\end{equation}
But, note that $A, B$, and $C$ define a Euclidean, coordinate triangle as well.  We apply the traditional law of cosines here:
\begin{equation}
|{\bf B} - {\bf C}|^2 = B^2 + C^2 - 2BC\cos(\alpha)
\label{lawofcos}
\end{equation}

Substituting (\ref{BandC}) and (\ref{lawofcos}) into (\ref{fin}) eliminates $B$ and $C$.  Clearing denominators and repeated application of the hyperbolic trigonometric identities yields the hyperbolic law of cosines:

\begin{equation}
\cosh(a) = \cosh(b) \cosh(c) - \sinh(b) \sinh(c) \cos(\alpha).
\end{equation}

Alternatively, there is a dual form of this relation that one can obtain from the cyclic permutations of the hyperbolic law of cosines together with more trigonometric identities (hyperbolic or otherwise):

\begin{equation}
\cos(\alpha) = \cos(\beta) \cos(\gamma) + \sin(\beta) \sin(\gamma) \cosh(a)
\end{equation} 

\subsection{Area Within a Given Distance of a Geodesic}

Consider a quadrangular region in the hyperbolic plane bordered on one side by a geodesic of length $d$, and having geodesics of length $r$ as the sides adjacent to the first.  Finally, define the fourth side as the curve of constant geodesic curvature such that each point along it is a distance $r$ from the length-$d$ geodesic \cite{strips}.  Recall the geodesic deviation equation:

\begin{equation}
\frac{\partial^2 \xi}{\partial s^2} = -K\xi
\end{equation}
where $\xi(s)$ is the separation between two geodesics as a function of the arclength parameter along each, $s$.  Consider the two sides of length $r$ in the context of this equation.  The initial conditions are $\xi(0) = d$ and $\frac{\partial \xi}{\partial s}(0) = 0$ because the initial curve of length $d$ is itself a geodesic and so first variations of its length vanish.  We find $\xi(s) = d \cosh(\sqrt{-K}s)$.  The area enclosed by our strip must then be given by:

\begin{equation}
A = \int_{0}^{r} ds\,\, \xi(s)
\end{equation}
or,
\begin{equation}
A = \frac{d \sinh(r \sqrt{-K})}{\sqrt{-K}}
\end{equation}
as required.


\begin{thebibliography}{1}
\bibitem{comment2} The boundary curvature correction in \cite{MS} is the inward pointing mean curvature of the codimension one boundary with respect to the entire manifold, as quoted in \cite{cdmrdk}.  In two dimensions this expression simplifies, as written.
\end{thebibliography}
\end{document}